\documentclass[superscriptaddress, 11pt, single-spaced,floatfix]{revtex4-1}

\usepackage{epsfig}
\usepackage{verbatim}	 
\usepackage{amsmath,amssymb}
\usepackage{calrsfs}
\usepackage{times}
\usepackage{amsmath}
\usepackage{psfrag}
\usepackage{bbold}
\usepackage[english]{babel}
\usepackage{graphicx}
\usepackage{braket}
\usepackage{natbib}
\usepackage{lipsum}
\usepackage{multirow,tabularx}

\usepackage{times}  
\usepackage{setspace}   
\usepackage[title]{appendix}
\usepackage{lineno}
\usepackage{xcolor}
\usepackage{xspace}
\usepackage{booktabs}
\usepackage{siunitx}
\usepackage[normalem]{ulem}
\usepackage{xr-hyper} 
\usepackage{xr}

\makeatletter
\newcommand*{\addFileDependency}[1]{
	\typeout{(#1)}
	\@addtofilelist{#1}
	\IfFileExists{#1}{}{\typeout{No file #1.}}
}
\makeatother

\usepackage{hyperref}
\usepackage{cleveref}
\definecolor{NavyBlue}{rgb}{0.1, 0.4, 0.8}
\hypersetup{breaklinks = true,
	colorlinks = true, 
	linkcolor = NavyBlue,
	urlcolor  = gray,
	citecolor = NavyBlue,
	anchorcolor = NavyBlue}

\newcommand{\wsb}{\texttt{r/wallstreeetbets}\xspace}
\newcommand{\wsbs}{\texttt{WSB}\xspace}
\newcommand{\gme}{\texttt{GME}\xspace}
\newcommand{\dfv}{\texttt{u/DeepFuckingValue}\xspace}

\newcommand{\R}[1]{{\small \href{https://reddit.com/r/#1}{\texttt{r/#1}}}}

\begin{document}
	\title{From Reddit to Wall Street: The role of committed minorities in financial collective action}
	
	\author{Lorenzo Lucchini}
	\affiliation{FBK - Fondazione Bruno Kessler, 38123 Trento, Italy}%
	
	\author{Luca Maria Aiello}
	\affiliation{IT University of Copenhagen, DK}%
	
	\author{Laura Alessandretti}
	\affiliation{Technical University of Denmark, DK-2800 Kgs. Lyngby, DK}%
	
	\author{\\Gianmarco De Francisci Morales}
	\affiliation{ISI Foundation, via Chisola 5, 10126 Torino, Italy}
	
	\author{Michele Starnini}
	\affiliation{ISI Foundation, via Chisola 5, 10126 Torino, Italy}
	
	\author{Andrea Baronchelli}
	\thanks{Corresponding author: abaronchelli@turing.ac.uk}
	\affiliation{City University of London, Department of Mathematics, London EC1V 0HB, UK}
	\affiliation{The Alan Turing Institute, British Library, 96 Euston Road, London NW12DB, UK}
	\affiliation{UCL Centre for Blockchain Technologies, University College London, London, UK}
	
	\begin{abstract}
		\textbf{In January 2021, retail investors coordinated on Reddit to target short selling activity by hedge funds on GameStop shares, causing a surge in the share price and triggering significant losses for the funds involved. Such an effective collective action was unprecedented in finance, and its dynamics remain unclear. Here, we analyse Reddit and financial data and rationalise the events based on recent findings describing how a small fraction of committed individuals may trigger behavioural cascades. First, we operationalise the concept of individual commitment in financial discussions. Second, we show that the increase of commitment within Reddit predated the initial surge in price. Third, we reveal that initial committed users occupied a central position in the network of Reddit conversations. Finally, we show that the social identity of the broader Reddit community grew as the collective action unfolded. These findings shed light on financial collective action, as several observers anticipate it will grow in importance.}
	\end{abstract}
	
	\maketitle


	\section*{Introduction}\label{sec:main}
	In January 2021, the GameStop shares traded on the New York Stock Exchange experienced a classic ``short squeeze''~\cite{asquith2005short,barberis2005survey}.
	As the price sharply jumped higher, traders who had bet that its price would fall (i.e., who ``shorted'' it) were forced to buy it in order to prevent even greater losses, thus further promoting the price rally~\cite{barberis2005survey,GameStopshortsqueezeWikipedia,cnbc_melvin_citron_bail_out}.
	Victims of the squeeze were professional hedge funds, and particularly Melvin Capital Management who lost 53\% of its investments for a total estimated $4.5$ billion USD~\cite{MelvinCapitalLost}.
	The short squeeze was initially and primarily triggered by users of the subreddit \R{wallstreetbets} (\wsbs), a popular Internet forum on the social news website Reddit, who managed to translate online discussions into an highly coordinated financial operation.
	
	These events garnered huge attention from the media, professionals, and financial authorities.
	Notably, the US  Treasury Secretary Janet Yellen convened a meeting of financial regulators including the heads of the Securities and Exchange Commission, Federal Reserve, Federal Reserve Bank of New York, and the Commodity Futures Trading Commission to examine the GameStop squeeze~\cite{yellen}.
	Cindicator Capital, a fund specialized in digital assets, published a hiring call for a sentiment trader with three years of active trading experience and having been a member of WallStreetBets for more than a year with karma---a Reddit measure of ``how much good the user has done'' for the community---of more than \num{1000}~\cite{hire_redditor}.
	Finally, the House Committee on Financial Services of the U.S. Congress held a hearing titled \emph{`Game Stopped? Who Wins and Loses When Short Sellers, Social Media, and Retail Investors Collide'} to discuss the events~\cite{DFVHearing}.
	They called as witness Reddit user Keith Gill, known as \dfv on \wsbs, who had a central role in triggering the collective action.
	At the hearing, committee members expressed concern with respect to gamification of investment 
	\footnote{https://www.bloomberg.com/news/articles/2021-03-17/democrats-eye-game-like-trading-apps-at-house-hearing-on-markets}, 
	encouraged by trading platform such as Robinhood, largely adopted by retail investors due to low commissions.
	However, how the coordination on \wsbs took place in the first place remains unclear, despite the importance of clarifying this mechanism in order to assess risks and devise regulations.
	
	In this paper, we analyse discussions on \wsbs from Nov 27, 2020, to Feb 3, 2021 and investigate how they translated into collective action before and during the squeeze. 
	Motivated by recent theoretical~\cite{xie2011social,niu2017impact} and experimental~\cite{centola2018experimental} evidence that minorities of committed individuals may mobilise large fractions of a population~\cite{granovetter1978threshold,schelling2006micromotives,xie2011social} even when they are extremely small~\cite{iacopini2021vanishing}, we investigate whether committed users on \wsbs had a role in triggering the collective action.
	To this aim,  we operationalise the commitment of a user as an exhibited proof that the user has financial stakes in the asset. 
	We consider specific labels on \num{128033} posts (called \textit{flairs}) and apply computer vision to classify \num{22277} screenshot posted as proofs, to identify a total of \num{36128} events of commitment by \num{30133} users. 
	We show that a sustained commitment activity systematically predates the increase of GameStop share returns,
	while simple measures of public attention towards the phenomenon cannot predict the share increase.
	Additionally, we also show that the success of the squeeze operation determines a growth of the social identity of \wsbs participants, despite the continuous flow of new users into the group.
	Finally, we find that users who committed early occupy a central position in the discussion network, as reconstructed by \wsbs posts and comments, during the weeks preceding the stock price surge, while more peripheral users show commitment only in the last phases of the saga.

	\begin{table*}[!b]
		\caption{Key events relevant to the \gme short squeeze.}
		\label{tab:events}
		\begin{center}
			\begin{tabular}{lllp{0.60\textwidth}}
				\hline
				& Event & Date & Description \\ \hline
				\emph{a} & \gme earnings & 2020-12-08 & \gme earning reports revealed a $257\%$ increase in e-commerce revenues\\
				\emph{b} & New board & 2021-01-11 & \gme announced a renewed Board of Directors, which included experts in e-commerce.\\
				\emph{c} & Citron prediction & 2021-01-19 & Citron Research, a popular stock commentary website, published a piece predicting the value of the \gme stock would decrease. Citron Research is managed by Andrew Edward Left, a financial analyst and renowned short seller.\\
				\emph{d} & Elon Musk's tweet & 2021-01-26 &  Business magnate Elon Musk tweeted ``Gamestonk!!'' along with a link to \wsbs.\\
				\hline
			\end{tabular}
		\end{center}
	\end{table*}
	

	\subsection*{The GameStop saga}
	
	GameStop (\gme) is a U.S. video game retailer which was at the center of the short squeeze in January 2021.
	The timeline of the events around to the squeeze is summarized in Table~\ref{tab:events}, and it unfolded as follows.
	In 2019, Reddit user \dfv entered a long position on \gme and started sharing regular updates in \wsbs.
	On October 27, 2020, Reddit user \texttt{u/Stonksflyingup} shared a video explaining how a short position held by Melvin Capital, a hedge fund, could be used to trigger a short squeeze.
	On January 11, 2021, \gme announced a renewed Board of Directors, which included experts in e-commerce.
	This move was widely regarded as positive for the company, and sparked some initial chatter on \wsbs.
	On January 19, Citron Research (an investment website focused on shorting stocks) released a prediction that \gme's price would decrease rapidly.
	On January 22, users of \wsbs initiated the short squeeze.
	By January 26, the stock price increased more than 600\%, and its trading was halted several times due to its high volatility.
	On that same date, business magnate Elon Musk tweeted ``Gamestonk!!'' along with a link to \wsbs.
	On January 28, \gme reached its all-time intra-day highest price, and more than 1 million of its shares were deemed failed-to-deliver, which sealed the success of the squeeze.
	A failure to deliver is the inability of a party to deliver a tradable asset, or meet a contractual obligation; a typical example is the failure to deliver shares as part of a short transaction.
	On January 28, the financial service company Robinhood, whose trading application was popular among \wsbs users, halted all the purchases of \gme stocks.
	On February 1 and 2, the stock price declined substantially.
	
	By the end of January 2021, Melvin Capital, which had heavily shorted GameStop, declared to have covered its short position (i.e., closed it by buying the underlying stock).
	As a result, it lost 30\% of its value since the start of 2021, and suffered a loss of 53\% of its investments, i.e., more than 4 billion USD.

	\subsection*{The \wsb ecosystem}
	
	Reddit is a public discussion website structured in an ever-growing set of independent \emph{subreddits} dedicated to a broad range of topics.
	Users can submit new \emph{posts} to any subreddit, and other users can add \emph{comments} to existing posts or comments, thus creating nested conversation threads.
	One such subreddit is \wsb (\wsbs), a forum for investors and traders on Reddit, which self-describes with the tagline ``Like 4chan with a Bloomberg terminal''.
	It is dedicated to high-risk trades involving derivative financial products (e.g., options and futures, often leveraged), and is thus not targeted to the beginner investor, but to somewhat experienced retail traders.
	Created in 2012, as of June 2021, it counts more than 10M subscribers (self proclaimed `degenerates', but also known as `autists', `retards', and `apes', depending on the type of information shared on the subreddit).
	As clear from its description so far, the \wsbs community is known for its profane and juvenile humor, and has a well defined identity reinforced also by the common use of jargon (e.g., `stonks' for stocks, `tendies' for profits, and `diamond hands' or `paper hands' for people that hold stocks through turbulent times or sell them at the first loss, respectively).
	The popularity of this forum has increased in recent years (since 2017 especially), possibly also due to the widespread adoption of no-commission brokers and mobile online trading platforms such as  \url{robinhood.com}.

	The topics of discussion on the forum are varied, but there are some common patterns of behaviors which are also described in the FAQ~\cite{wsb_faq}.
	When submitting a post, a user can apply a category tag called `flair', which serves as an indication of its content.
	The allowed flairs, together with a short description, are reported in Table~\ref{tab:flair}.
	The community takes flairs seriously and strictly enforces them (e.g., the FAQ report that misusing important flairs can lead to getting permanently banned).
	It is thus very common to find posts containing screenshots of an open position on a risky bet tagged with a YOLO flair, all interspersed with unhinged humorous posts and memes.
	
	The discussion within the subreddit follows a simple post-comment dynamic, where each post separately grows its multi-level comment tree.
	Each interaction, it being a post or a comment, can additionally receive `upvotes' and `downvotes'.
	While `upvoting'/ or `downvoting' represents a typical `slacktivist' practice for anonymously expressing one's position, other users can also choose to `award' prizes to more emphatically recognize a post or comment.

	\begin{table*}
		\caption{Flairs allowed on \wsb and their meaning as per the subreddit guidelines.}
		\label{tab:flair}
		\begin{center}
			\begin{tabular}{l p{0.65\textwidth}}
				\hline
				Flair & Meaning \\ \hline
				YOLO (You Only Live Once) & YOLO flair is for dank trades only. The minimum value at risk must be at least \$10,000 in options, or \$25,000 in equity.\\
				DD (Due Diligence) & The research you have done on a specific company/sector/trade idea. This is a high effort text post. It should include sources and citations. It should be a long post and not just a link to a submission.\\
				Discussion & An idea or article that you would like to talk about. Needs to be more involved than "up or down today?"\\
				Gain & Use this flair to show off a solid winning trade. Minimum gain is \$2,500 for options, \$10,000 for shares. You must show or explain your trade. If you have to say something like "position in comments" then it's a bad screenshot.\\
				Loss & Show off a brutal, crushing loss. Minimum loss \$2,500 for options, \$10,000 for shares. You must show or explain your trade. If you have to say something like "position in comments" then it's a bad screenshot.\\ \hline
			\end{tabular}
		\end{center}
	\end{table*}

	\section*{Results}\label{sec:results}
	\subsection*{Collective attention, Commitment, and Identity on WSB}
	
	\begin{figure}
		\centering
		\includegraphics[width=0.45\textwidth]{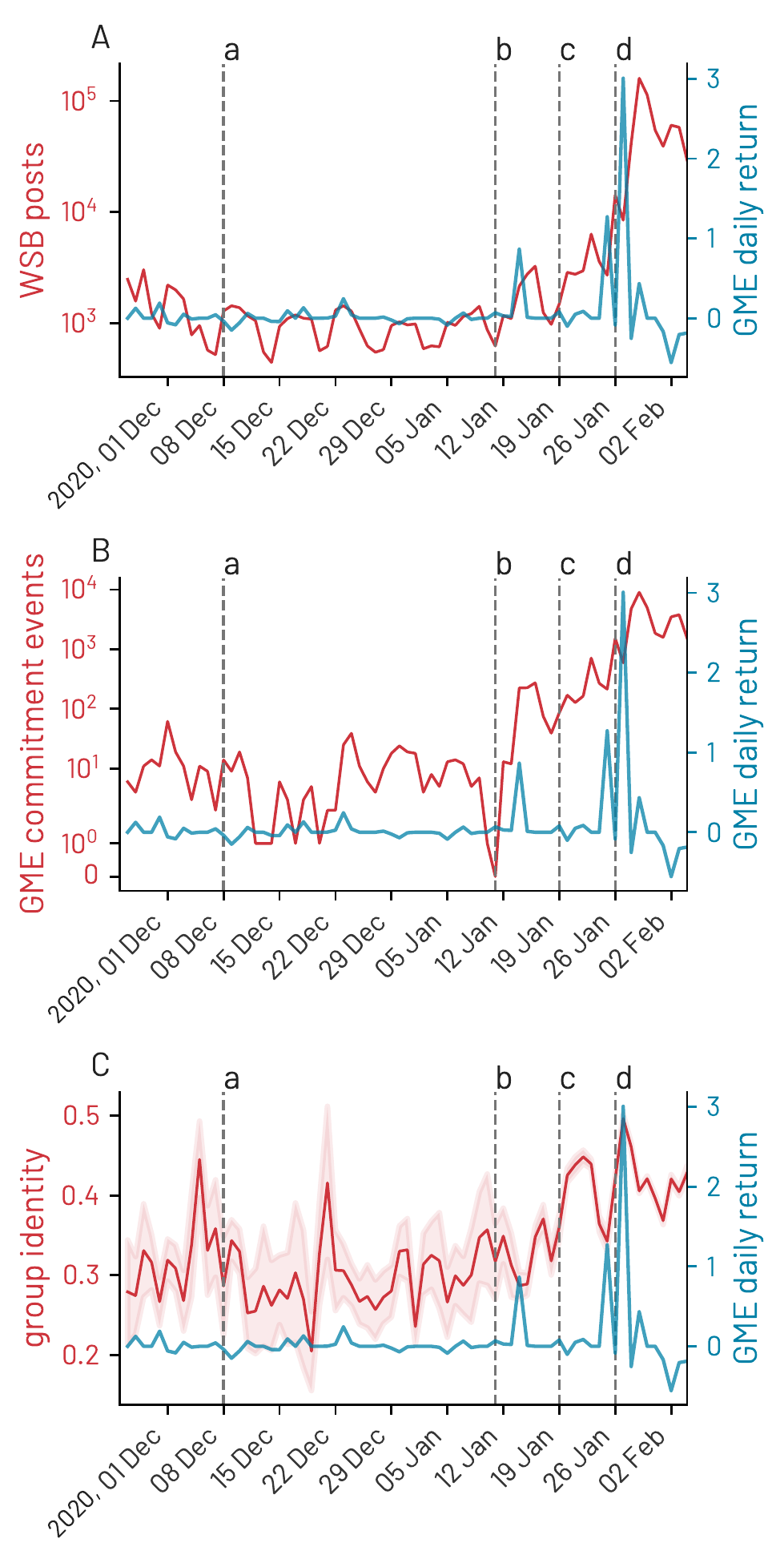}
		\caption{\gme stock returns compared with: (a) number of posts submitted on \wsbs; (b) number of posts on \wsbs that showed financial commitment; (c) level of group identity (shaded areas corresponds to two standard errors of the daily average).}
		\label{Figure 1}
	\end{figure}
	
	Figure~\ref{Figure 1} compares the daily returns of \gme (the percent increase of price compared to the previous day) with three quantities calculated over time on a daily basis: \emph{i)} activity within the community measured as the number of posts submitted on \wsbs; \emph{ii)} number of posts on \wsbs that showed financial commitment towards \gme stocks; and \emph{iii)} level of group identity signaled by language markers in \wsbs submissions about \gme.
	
	The posting activity in the \wsbs community is characterised by a weekly periodicity that endures stably until the announcement of a renewed Board of Directors (event \emph{b}).
	After the first considerable increase in the stock price on January 14 ($+57$\%), the activity grows noticeably.
	Posting activity raises exponentially after the second price spike on January 26 ($+92$\%) and it culminates on January 28, two days after the stock evaluation reached its maximum.
	Public attention to the \gme phenomenon spreads far beyond the boundaries of Reddit.
	The number of \gme-related tweets follows a similar exponential growth starting on January 27 after Elon Musk's endorsement (event \emph{d}) and peaking on January 28 (Figure SI~\ref{fig:SIcollectiveaction}(a)).
	The growing interest on Twitter matches the explosive growth in the number of new subscribers to the \wsbs subreddit (Figure SI~\ref{fig:SIcollectiveaction}(b-c)).
	Overall, these results support three conclusions.
	First, collective attention towards \gme follows the asset price growth with a delay.
	Second, despite the collective action being designed and coordinated on Reddit, wide interest was expressed on other social media as well.
	Last, not only the discussion originated from Reddit gradually attracted the attention of larger crowds to the topic, but it also engaged those crowds to the point of attracting them to the original source of the discussion---the \wsbs subreddit.

	The evolution of commitment over time differs considerably from the growth of collective attention.
	Figure~\ref{Figure 1}(b) shows the number of daily commitment events measured by counting ``Gain'', ``Loss'', ``YOLO'' posts (i.e., posts with one of these flairs), and the screenshots that \wsbs users submitted as proof of stake (see Methods for details).
	Before the new board of directors was announced (event \emph{b}), \wsbs users uploaded a few dozens of commitment posts per day.
	The number of commitment posts increases tenfold on the day of the first price spike and keeps growing steadily afterwards.
	For eleven days, between the first price spike on January 14 until the next spike on January 25, such increase in commitment takes place in absence of any growth in financial returns.
	In summary, commitment predates price surges and is sustained also in absence of gains.
	
	The presence of commitment in absence of returns raises the question of whether commitment was supported by other processes endogenous to the \wsbs community.
	A recent ethnographic study found that active members of \wsbs use shared linguistic markers and reciprocation of custom awards to express and reinforce the community's sense of identity~\cite{boylston2021wallstreetbets}.
	Identity is a shared sense of belonging to a group~\cite{brewer1996we} that can influence inter-group behavior~\cite{tajfel1979integrative}, not least by fostering cooperation~\cite{yamagishi2000group,simpson2006social}.
	Identity is often signaled explicitly through symbols~\cite{mach1993symbols} or language cues~\cite{ochs1993constructing}.
	To measure the group identity in \gme-related submissions, we used a validated indicator of group identity~\cite{tausczik2010psychological}: for each submission, we calculated the fraction of the first person pronouns that are the plural pronoun \emph{``we''}, and averaged those fractions across all the submissions of a given day (more details in Methods).
	Figure~\ref{Figure 1}(c) shows the group identity expressed by \gme-related submissions within the \wsbs community.
	The signal oscillates heavily until mid-January, due to the relatively low number of submissions.
	As the number of submissions increases, we detect two peaks.
	The first peak follows the market analysis from Citron Research (event \emph{c}) that forecast a drop in \gme stock price and antagonized the members of the \wsbs community by referring to them as ``suckers at this poker game''.
	This finding is in agreement with the theoretical expectation of community identity being created during processes of struggle between social groups~\cite{cook2008role}---in this case, between \wsbs and its detractors.
	The second peak matches the maximum increase of the stock price, and it is likely caused by the acknowledgment of collective success in performing the short squeeze.
	In short, we find that expressions of identity emerged concurrently with the increase of commitment and might have played a role in sustaining it, but identity is unlikely to be the origin of the collective action.

	\subsection*{Commitment and reach of core vs peripheral authors}
	
\begin{figure*}
		\centering
		\includegraphics[width=.9\textwidth]{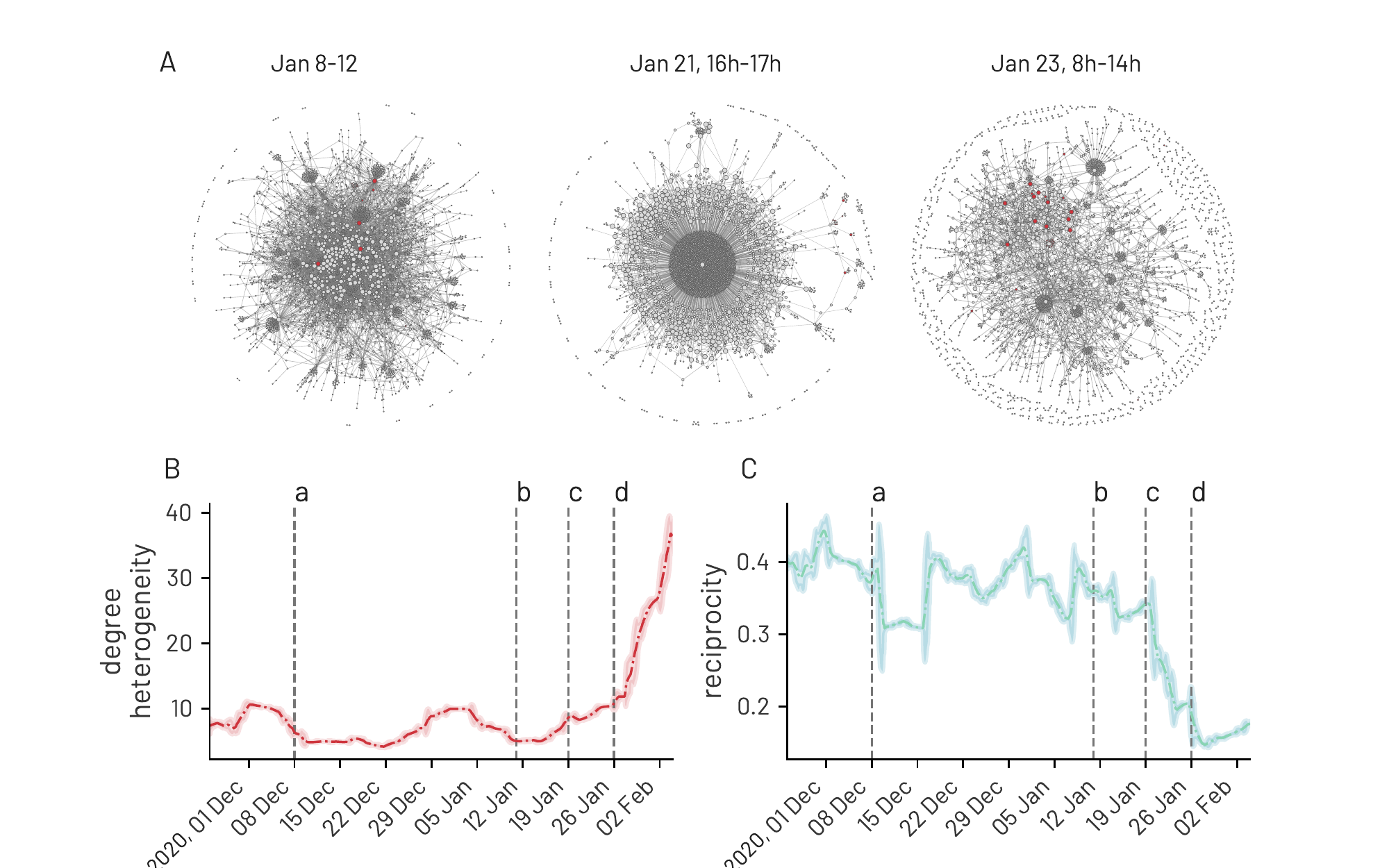}
		\caption{\emph{The evolution of the \gme discussion network.} 
			Panel (a) shows three examples of \gme discussion networks, reconstructed in different time windows with the same number of nodes, $N$ = 3000.
			Nodes are \wsbs users, colored according to whether they posted a commitment submission (red) or not (gray). The size of nodes is inversely proportional to their k-shell, i.e. nodes belonging to the core are bigger than peripheral nodes. 
			A link exists between two nodes if one of the two replied to the other at least once.
			Panels (b) and (c) show two key topological features of networks reconstructed over a rolling time window of 7 days:
			(b) The heterogeneity of the degree distribution $\kappa$, defined as $\kappa \equiv \langle k^2 \rangle / \langle k \rangle^2$, where $\langle k \rangle$ and $\langle k^2 \rangle$ are the first and second moments of the degree distribution, and \emph{n}-th moment is $\langle k^n \rangle \equiv \sum_i^N k_i^n/N$~\cite{Newman2010}. 
			(c) The average network reciprocity. Shaded areas represent $2$ standard deviations of the network metrics aggregated on a daily rolling window basis.}
		\label{fig:networks}
	\end{figure*}

	The sustained flow of commitment events during the weeks preceding stock price surge indicates the presence of a minority of committed users.
	As the interaction between the committed minority and the rest of the community is crucial to the success of a collective action~\cite{granovetter1978threshold,schelling2006micromotives,xie2011social}, we study the dynamics of social interactions between committed individuals and other \wsbs users.
	These interactions occur over a rapidly evolving social network.
	Figure~\ref{fig:networks}(a) shows a few snapshots of the network of replies over time.
	In these networks, users are connected if they submitted a comment in reply to the post or comment of another during the time-span considered.
	As new users join, the number of small disconnected components in the network increases (see Figure SI~\ref{fig:SInetworkevolution}(b-c)), while the connected component tends to cluster around few popular discussion threads (especially the so-called \emph{daily megathreads}~\cite{boylston2021wallstreetbets}) created with the purpose of summarizing the events of the day and planning future actions.
	The structural transformation of the network happens abruptly rather than gradually.

	We quantify this structural change by reconstructing networks over a rolling time window of 7 days, and looking at the evolution of two key topological quantities of these networks in time.
	First, we observe that the heterogeneity of the distribution of the nodes' degree (i.e., the number of different users each user replies to) \cite{Newman2010} increases three-fold in the span of 20 days after event \emph{c}, thus reflecting the simultaneous emergence of super-hubs of discussion together with users engaging only in isolated interactions. 
	Second, the direct reciprocity of interaction (i.e., the fraction of replies that are reciprocated within the time window considered) gets roughly halved in the same time span.
	This signal, combined with the increase in expressions of group identity (Figure~\ref{Figure 1}(c)), is compatible with the emergence of generalized reciprocity~\cite{yamagishi2000group}, a norm according to which individual messages are not expected to receive direct responses; comments are not perceived as pieces of a conversation but rather as contributions to a collective discussion from which everyone benefits.

	The complex and dynamic nature of the social network raises the question of what is the typical position of committed users in the network, and whether this position changes over time.
	To answer this question, we operationalise the notion of network position with the concept of $k$\emph{-core shell}~\cite{seidman1983network}: the set of nodes in which every node is connected with other members of the set with at least $k$ links.
	This measure is a good indicator of a node's centrality because it directly gauges embeddedness (the density of connections around it), and it is a good proxy for reachability (how quickly it can be reached from any other node of the network).
	For each temporal slice of the network, we perform its $k$-core decomposition (see Methods), and we measure the level of commitment exhibited in each $k$-core shell.
	Borrowing from previous work~\cite{barbera2015critical}, we estimate the potential influence that commitment events have on the community at large by measuring not only the volume of commitment events in a shell, but also the number of people that these events \emph{reach}---namely the number of \wsbs members who commented on a post that is submitted by a committed node in that shell. 
	
\begin{figure*}
		\centering
		\includegraphics[width=\textwidth]{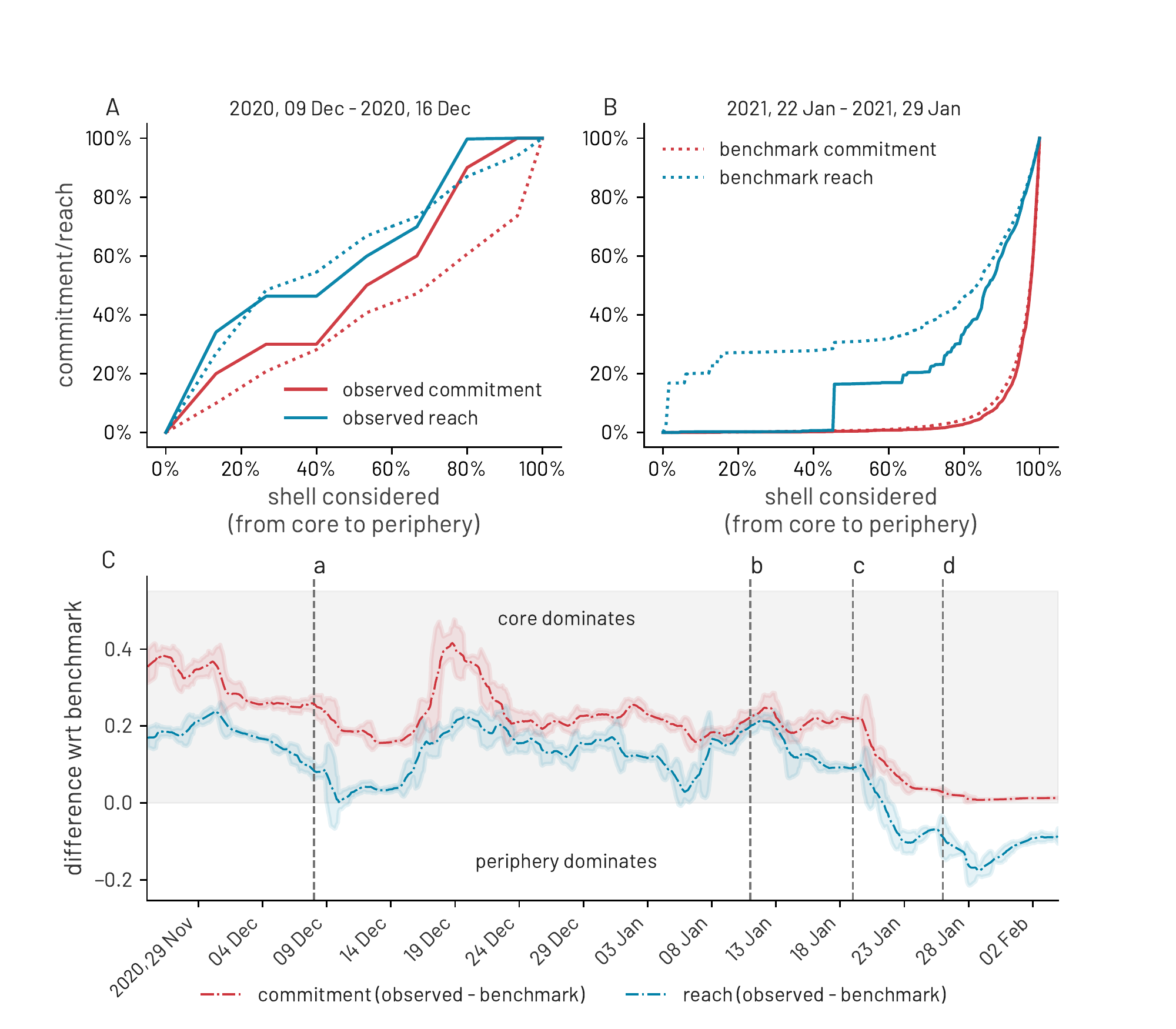}
		\caption{\emph{The evolution of commitment and reach.} (a-b) Fraction of observed commitment activity (red line) and reach (blue line) produced by nodes belonging to an increasingly large fraction of $k$-core shells (from core to periphery). Curves constructed using the observed data (filled lines) are compared to those obtained for the benchmark model (dashed lines). Results are shown for two network slices, the first constructed between December 11 and December 18 (a), the second between January 20 and January 27 (b). (c) Average difference between the area below the observed curve and the area below the benchmark curve over time, for commitment (red dashed-dotted line) and reach (blue dashed-dotted line). Shaded areas corresponds to the $2$ standard deviation area computed for each slice. Dashed vertical lines indicate relevant events (see Table \ref{tab:events}). For values of difference larger than zero (gray shaded area), activity is concentrated in the core of the network, relative to the benchmark model. Networks are constructed using a sliding window of $7$ days.
		}
		\label{fig:reach_activity}
	\end{figure*}

	Figure~\ref{fig:reach_activity} shows how commitment activity and its reach are distributed between users in the core of the network (high $k$-core shells) and peripheral users (low $k$-core shells), as a function of time.
	First, in Figure~\ref{fig:reach_activity}(a-b) we show the fraction of commitment activity and reach that are generated by nodes belonging to an increasingly large number of $k$-core shells, taken from the core to the periphery.
	To disentangle the effect of the network's evolution from the distribution of commitment and reach on the network, and meaningfully compare commitment distribution over networks reconstructed in different periods, we contrast the observed commitment activity and its reach to a null model benchmark which preserves the commitment but randomizes the network's topology (see Methods).
	In Figure~\ref{fig:reach_activity}(a-b), the curve being higher (lower) than the benchmark indicates that commitment volume or reach are generated predominantly by nodes in the core (periphery) of the network.
	For example, in the network of interactions between November 11 and December 18, central nodes are those who pledge more commitment to the \gme cause (Figure~\ref{fig:reach_activity}(a)); on the contrary, when considering interactions between January 20 and January 27, commitment comes mostly from peripheral actors (Figure~\ref{fig:reach_activity}(b)).
	
	To get a comprehensive picture of the coreness of committed users over time, we measure the difference between the area below the observed curve and the area below the benchmark curve at a given temporal slice, for all the slices computed on networks reconstructed by using a rolling time window of 7 days.
	Results are robust to the slicing strategy chosen for constructing the networks (see Supplementary Information (SI), Sec.~\ref{sec:SIslicingRobustness}, Fig. SI~\ref{fig:SInodeSlicing} and \ref{fig:SIsubsSlicing}).
	Figure~\ref{fig:reach_activity}(c) shows the value of such difference as a function of time. Relative to the benchmark model, both commitment and reach are concentrated in the network's core until event \emph{c} (January 19).
	From that moment onward, the commitment activity obtains a larger reach within the periphery.
	While always remaining more concentrated in the core, the commitment activity spreads more and more towards the periphery following event \emph{c}.
	Therefore, the committed minority which may have triggered the first price increase in the \gme stock is formed by central users in the discussion network unfolding on \wsbs.
	Only in the last phases of the collective action, when the price has already increased considerably, peripheral users step in and show commitment, which reaches more peripheral peers.

	\section*{Discussion}\label{sec:discussion}
	In this paper we showed that the collective action originated on Reddit and culminated in the successful short squeeze of GameStop shares was driven by a small number of committed individuals.
	We operationalised financial commitment on Reddit as providing proof of stakes in a given asset, often in the form of a screenshot.
	We then showed that events of commitment predated the initial surge in price, which in turn attracted more participants to the GameStop discussion and thus triggered new events of commitment.
	Finally, we described how initial committed users were part of the core of the network of Reddit conversations, and that the social identity of the broader group of Reddit users grew as the collective action unfolds. 
	
	Our study focused on an single, unprecedented, event of financial collective action.
	While this is certainly a limitation, as more events would allow us to corroborate or falsify our findings, a prompt investigation of the GameStop events was in order.
	The events that unfolded over the course of the few weeks that we analysed in our study caused sustained effects on the market.
	Seven months later (at the time of writing this manuscript), the value of the GameStop stocks had risen by $1000\%$ compared to the beginning of 2021.
	The price increase inflicted enormous financial losses to multiple hedge funds, one of which was forced to shut down~\cite{fletcher21hedge}.
	
	The influence of retail investors in equity markets is rapidly growing, and now accounts for almost as much volume as hedge and mutual funds combined.\footnote{https://www.ft.com/content/7a91e3ea-b9ec-4611-9a03-a8dd3b8bddb5}
	This rise has been mainly driven by the emergence of commission-free trading platforms, that offer the possibility to trade fractions of shares, so that users can start trading even with very small amounts. 
	Moreover, these platforms allow investors to use leverage, by buying and selling options and accessing to cheap margin loans from brokerages, in a gamified user experience.
	This ``democratization of trading and investing" is unlikely to disappear any time soon~\cite{aramonte2021rising}, so other financial collective actions might be coordinated in the future, possibly through different social media channels.
	
	In this perspective, beyond the role of committed individuals in promoting the coordinated action, our findings have other potential implications to be tested in future research. 
	(i) The fact that initial committed individuals were part of the core of the Reddit discussions implies that the system may be resilient against adversarial attacks where freshly created ``committed'' bots try to influence the community. 
	(ii) The finding that identity was not the driver of the collective action but, on the contrary, a byproduct of it may imply that successive actions that leverage it might be easier to coordinate. 
	(iii) The change in network structure ensuing from the arrival of new users, who joined the discussion motivated by the initial success of the squeeze, and the corresponding shift of the bulk of commitment and reach from the core to the periphery of the network, highlights the role of the system's openness and the hierarchies that catalyse a successful collective action.
	
	Taken together, our findings highlight that financial collective action cannot be reduced to the impact of social coordination on financial markets.
	The effect --and, particularly, the success-- of an action have profound consequences on the membership, structure, and dynamics of the original group, whose evolution may have in its turn consequences on future actions.
	Thus, the initial committed individuals trigger a behavioural cascade which is self-sustaining and transforms the group itself.
	More events and data are needed to clarify this interplay between bottom-up processes of social coordination and financial markets, and this is a direction for future work.
	Our results represent a first step in this direction, and we anticipate that, as financial collective action is expected to acquire even more importance in the future, they will be of interest to researchers, industry professionals, and regulators.
	
	\section*{Methods}\label{sec:methods}
	\subsection*{Data}
	We used two main sources of data: the activity on the subreddit \R{wallstreetbets} and the price of GameStop shares, ticker \gme.

	Reddit is organized in communities, called \emph{subreddits}, that share a common topic and a specific set of rules.
	Users subscribe to subreddits, which contribute to the news feed of the user (their home) with new posts. Inside each subreddit, a user can publish \emph{posts} (also called ``submissions''), or \emph{comment} on other posts and comments, thus creating trees of discussion that grow over time.
	Users can attach \emph{flairs} to posts: a set of community-defined tags to define the semantic scope of the post, thus  facilitating  content search and filtering.
	Users can assign \emph{awards} to posts or comments to recognize their value.
	Awards are sold by Reddit for money, they come in a variety of types, and some of them reward the recipient with money or perks such as access to exclusive subreddits.
	
	We collected all posts and comments submitted to the \R{wallstreetbets} subreddit from January $1$, $2016$ up to the beginning of February $2021$. We did so by querying the \textit{Pushshift API}~\cite{baumgartner2020pushshift}, which stores all Reddit activity over time---using the \textit{PMAW} wrapper~\cite{pmaw_github} (see SI, Sec~\ref{subsec:SIdataretrieval} for more details).
	The API returns rich metadata, including the timestamp of submission, the identity of the authors, its text content, and the awards each submission and comment received. 
	In total we retrieved \num{1132897} posts and \num{29566180} comments submitted to the subreddit by $1\,364\,080$ different authors.
	We specialize only to posts related to GME by searching for posts containing either in the title of in the text-body the word ``GME'' or ``Gamestop'' (lowercase occurrences included) and all the comment trees associated with those submissions.
	This selected set consists of \num{129731} posts and \num{2575742} comments.
	The period over which our study focuses its attention (from November 27, 2020 to February 3, 2021) includes $99\%$ of the posts and $98\%$ of the comments submitted since January 1, 2016 until February 3, 2021.

	We retrieved GameStop daily prices from \textit{Yahoo Finance}, using the Python library \textit{yfinance}~\cite{yfinance}, and computed the daily price return as the daily relative change, $r(t) = \frac{p(t)}{p(t-1)} -1$ where $p(t)$ is the \textit{Open} price at day $t$.
	
	\subsection*{Quantifying commitment} \label{subsec:commitment}
	
	\begin{table*}[tbp]
		\caption{Commitment events per type. Count column shows the number of posts classified as commitment events because of a ``YOLO'', ``Gain'', ``Loss' flair, or a screenshot of commitment identified with our machine vision classifier. Unique count shows the number of posts uniquely classified by the commitment type.}
		\begin{center}
			\label{tab:commitmentTypes}
			\begin{tabular}{l p{0.10\textwidth} p{0.15\textwidth} p{0.15\textwidth} p{0.15\textwidth}}
				\hline
				Event type & Count & Unique count & Authors & Unique authors\\
				\hline
				YOLO & $23\,230$ & $21\,455$ & $20\,107$& $18\,484$\\
				Gain & $7\,293$& $6\,339$ & $6\,160$& $5\,293$\\
				Loss & $2\,986$& $2\,387$ & $2\,768$& $2\,198$\\
				Pictures & $5\,947$ & $2\,619$ & $5\,326$& $2\,478$\\
				\hline
			\end{tabular}
		\end{center}
	\end{table*}

	One of the widely-shared norms in the \wsbs community is to provide proof of one own's financial position when initiating a new discussion about investments~\cite{boylston2021wallstreetbets}. 
	This is commonly achieved by supplementing submissions with screenshots of open positions---typically gains, losses, or orders---taken from online trading applications.
	We used these screenshots to quantify commitment, as they provide a direct way to identify users who had stakes in financial assets.
	To gather them, we employ two methods: \emph{flairs} and screenshots. 
	
	We use three flairs to mark posts containing a proof of position: the \emph{gain} and \emph{loss} flairs mark gains or losses for a minimum of \num{2500} USD, and the \emph{YOLO} flair indicates investment positions with a minimum value at risk of \num{10000} USD. 
	Flair-tagged submissions are moderated and are approved only if a relevant screenshot is attached. 
	While it is mandatory for users to attach investment screenshots to have flairs approved, they can also attach screenshots to their submissions without using any flair.
	
	As we are interested in capturing any signal of commitment, regardless of their magnitude, we resort to machine vision to identify commitment screenshots based on their visual content only.
	We retrieve all the screenshots attached to any of the submissions in our dataset by querying all URLs terminating with common image extensions (e.g., .png, .jpg).
	Out of this set, we randomly sample \num{3745} images and manually inspect them.
	We mark as \emph{positive} all the screenshots which display gains, losses, or orders, and as \emph{negative} all the remaining images, which include a broad variety of content ranging from screenshots of stock prices to memes.
	
	We label \num{1042} positive examples and \num{2703} negative examples.
	We use this set of labeled images to train a supervised model.
	Among several classifiers available off-the-shelf that we test, the most accurate is a PyTorch~\cite{paszke2019pytorch} implementation of DenseNet~\cite{huang2017densely}, a deep neural network architecture designed for image classification.
	We initialize DenseNet with weights pre-trained on ImageNet~\cite{deng2009imagenet}, a widely-used reference dataset of 1.2M labeled images.
	We then fine-tune the neural network (i.e., update its weights) by training it further by feeding it 70\% of our labeled images.
	During fine-tuning, we use the Adam optimizer~\cite{kingma2014adam} to minimize cross-entropy loss.
	We then measure the classifier's performance on the remaining 30\% of the examples by using \emph{precision} (the fraction of pictures that the classifier labeled as positive that are actually positive), \emph{recall} (the fraction of positive pictures that the classifier labeled correctly), and F1 score (the harmonic mean between precision and recall).
	On our validation set, the classifier achieves a precision of $0.85$, a recall of $0.73$, and an F1 score of $0.77$. 
	
	We run the classifier on all images from \wsb and we merge the posts which contain the images that the classifier marks as positive with the set of flaired posts. 
	In total, following this procedure we identify $3\,6128$ commitment events.
	Table~\ref{tab:commitmentTypes} shows the number of commitment events divided by event type. Posts can be classified as commitment events by flair type or pictures of a holding position or both at the same time.
	The ``unique count'' column shows the contribution to the identification of commitment events uniquely coming from the single commitment type.
	In Figure SI~\ref{fig:SIcommitmentcomposition} we show the contributions of each commitment, revealing that commitments from ``YOLO'' flairs are the dominant ones except for a three-days period in which the first price surge was followed by a surge in the number of commitment events from ``Gain'' flairs.
	
	\subsection*{Quantifying identity}
	
	To capture linguistic expression of identity, we use two methods.
	First, we resort to a simple word count approach using Linguistic Inquiry Word Count (LIWC).
	LIWC is a lexicon of words grouped into categories that reflect social processes, emotions, and basic functions.
	It is based on the premise that the words people use provide clues to their psychological states.
	In particular, the abundant use of words in the LIWC category \emph{we} (i.e., first-person plural subject pronoun) related to the use of words from the LIWC category \emph{I} (i.e., first-person singular subject pronoun) is a validated indicator of group identity~\cite{tausczik2010psychological}.
	Therefore, we measure identity as the fraction of pronoun \emph{we} against the number of both \emph{we} and \emph{I} pronouns occurring in each submission text body.
	The results obtained with this particular estimator of identity are robust when compared to two alternative methods, which we discuss in SI.

	\subsection*{Discussion network on WSB}
	
	
	We reconstructed the network of social interactions on \wsb; each node represents a user who submits a post or a comment on the subreddit, and each directed link $(u,v)$ represents user $u$ commenting on a submission by user $v$.
	The direction of the link represents an interaction, and it is opposite to the information flow (user $u$ should have read what $v$ wrote to answer, but it is not guaranteed that $v$ will read $u$'s reply).
	Considering all nodes and interactions on \wsbs between authors discussing about \gme, the resulting networked components consist of $N=\num{450710}$ nodes and $E=\num{2649108}$ directed links, activated over the entire period starting on November 27, 2020 and concluding on February 3, 2021.
	
	The time at which posts and comments are published can be used to obtain a description of social interaction dynamics. We modeled such dynamic through temporal \emph{slicing}.
	In particular, we considered a rolling time window of seven days and shift it by two hours throughout the whole timespan of our dataset, for a total of \num{1092} windows.
	For each time window, we constructed a network using posts and comments published during that time window.
	We tested alternative temporal slicing strategies, and discussed them in SI Sec.~\ref{sec:SIslicingRobustness}.
	
	For each slice, we characterized nodes with a number of features, including their age (the time elapsed since their first interaction within the community), in- or out-degree (number of incoming or outgoing edges), their commitment (number of commitment events), or the reach of their commitment (number of users who comment on their commitment events). We also ran $k$-core decomposition~\cite{wasserman1994social} on the network of each temporal slice.
	The algorithm partitions nodes by their \emph{core shell} (or core number), i.e., the shell $k$, defined as the maximal subgraph in which every vertex has at least degree $k$.
	The $k$-core decomposition algorithm does not take edge directionality into account, and it considers the degree of a node as the sum of its outgoing and incoming edges.
	
	\smallskip\noindent
	\textbf{Null model for random commitment activity.}
	When computing the commitment of nodes as a function of their core number, to assess if committed users are more central or peripheral in the network, it is important to compare with a null model which takes into account the network's topology. 
	For this reason, we consider a null model of random commitment in which committed events are reshuffled randomly over the whole network, while the network's structure is preserved. 
	The empirical commitment of nodes with core number $k$ is then compared with a uniform distribution of commitment across nodes, which is equivalent to averaging the results over an infinite number of random shuffles.

	
	\bibliographystyle{unsrt}
	\bibliography{bibliography}

	\section*{Author contributions}
	All authors conceived and designed the experiments. L.L., L.M.A., L.A., and G.D.F.M. performed the experiments. All authors contributed in analyzing the data and writing the manuscript. 
	
	\section*{Competing interests}
	All authors declare no competing interests. 
	
	\cleardoublepage
	\setcounter{figure}{0}
	\renewcommand{\figurename}{Figure SI}
	
	\widetext

	\appendix
	\renewcommand{\thesubsection}{A.\arabic{subsection}}
	\begin{center}
		\section*{\Large{\textbf{Supplementary Information}}}
	\end{center}
	
	\vspace{1cm}
	
	\subsection{Data retrieval}\label{subsec:SIdataretrieval}
	
	We downloaded Reddit data mostly from Pushshift, an API which regularly copies Reddit activity data directly from the Reddit API~\cite{baumgartner2020pushshift}. However, on January 25 and until January 27 the Pushshift servers interrupted from retrieving Reddit comments. In our analysis we filled this gap in the data by calling the Reddit API using the python wrapper PRAW~\cite{PRAWThePythonRedditAPIWrapper}.
	Specifically, we requested the comment trees for \num{234881} posts and successfully retrieved the information for \num{234866} of them.
	The missing $15$ comment trees required an ad-hoc procedure because of their size.
	We directly requested comments for such submission by scraping the json version of their Reddit page and iterating until the tree was completely reconstructed.
	The additional comments retrieved following this two steps procedure are $322\,773$.

	\subsection{Measures of collective attention}\label{subsec:SI_twitter_wsb_subscribers}
\begin{figure}
		\centering
		\includegraphics[width=0.45\textwidth]{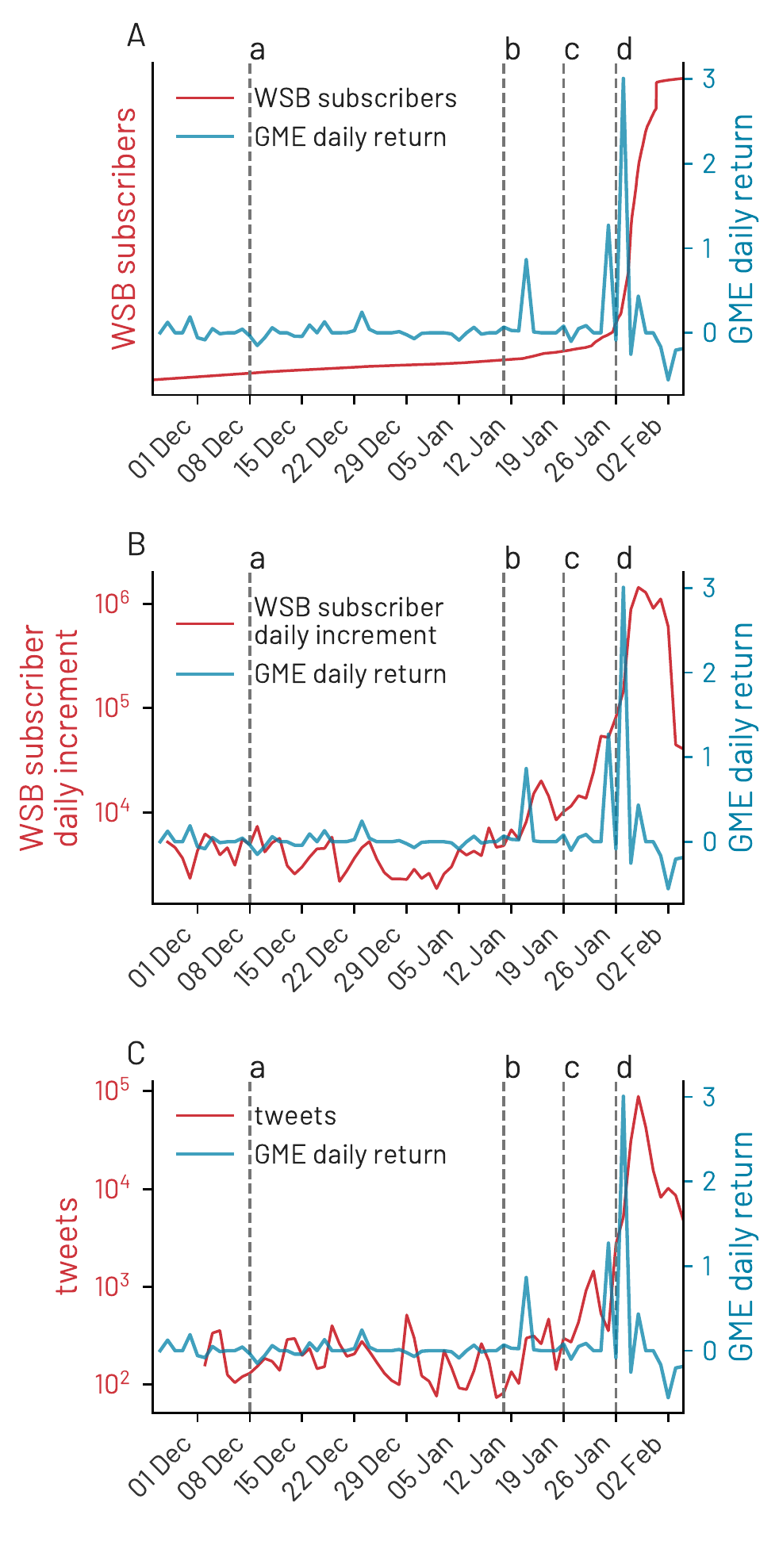}
		\caption{\emph{Measures of collective attention.} We compare the daily returns (blue) with (a) total number of \wsbs subscribers in time, (b) daily relative increment in subscriber number, and (c) Twitter activity, measured as the volume of  \gme-related tweets.}
		\label{fig:SIcollectiveaction}
	\end{figure}

	Contrary to the sustained trend of commitment to the \gme stock, which was present days before the asset price grew, the involvement of large number of people in the community occurred only weeks after the first signals of commitment. To test the robustness of this observation, we complemented the activity measure presented in Figure~\ref{Figure 1}A with the number of WSB subreddit subscribers, and its daily relative increment (see Figure~\ref{fig:SIcollectiveaction} (a) and (b)). Additionally, to measure exogenous interest towards \gme, we gathered data from Twitter. We queried the Twitter API for academic research (which allows to retrieve tweets at any point in the past) to download tweets with the \#GME or \#Gamestop hashtags posted between December 1st, 2020 and February 5th, 2021. We collected a total of \num{5640149} tweets, whose volume over time is shown in Figure~\ref{fig:SIcollectiveaction}C. All indicators support the observation that commitment predates not only the surge in prices but also the the mass attention that \gme received as a result of the short squeeze success.
	
	\subsection{Measures of commitment events}\label{subsec:SIcommitment}
\begin{figure}
		\centering
		\includegraphics[width=.45\textwidth]{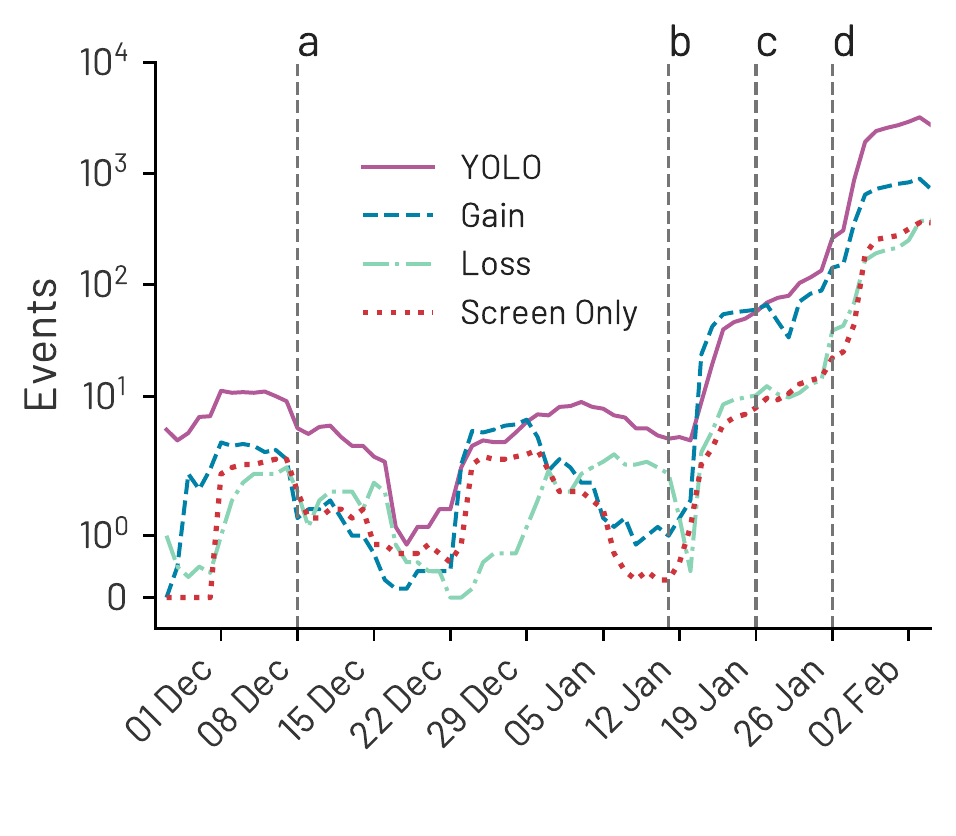}
		\caption{\emph{Composition of commitment events  by flair and proof of ownership.} The figure  shows how the commitment events are distributed in time over the period of interest. The number of commitment events are counted on a daily basis. Commitment events are dominated for most of the time by ``YOLO'' posts.}
		\label{fig:SIcommitmentcomposition}
	\end{figure}

	As discussed in the manuscript, we identified \gme commitment events with a two-step procedure. First we selected \gme-related event by looking for submissions whose title or text (the \textit{selftext} field returned from the Pushshift API) contains the token ``GameStop'' or the ticker name ``\gme'' (both either in uppercase or in lowercase letters). Then, we used the \wsb subreddit flair classification system identifying as commitment events those \gme-related submissions with either ``YOLO'', ``Gain'', or ``Loss'' flair. Additionally, we also include submissions sharing a screenshot of a proof of ownership estimated using computer vision. In Figure~\ref{fig:SIcommitmentcomposition}, we show the composition of commitment events for the four different categories.
	
	\subsection{Measures of group identity}
	
	\begin{figure*}
		\centering
		\includegraphics[width=.9\linewidth]{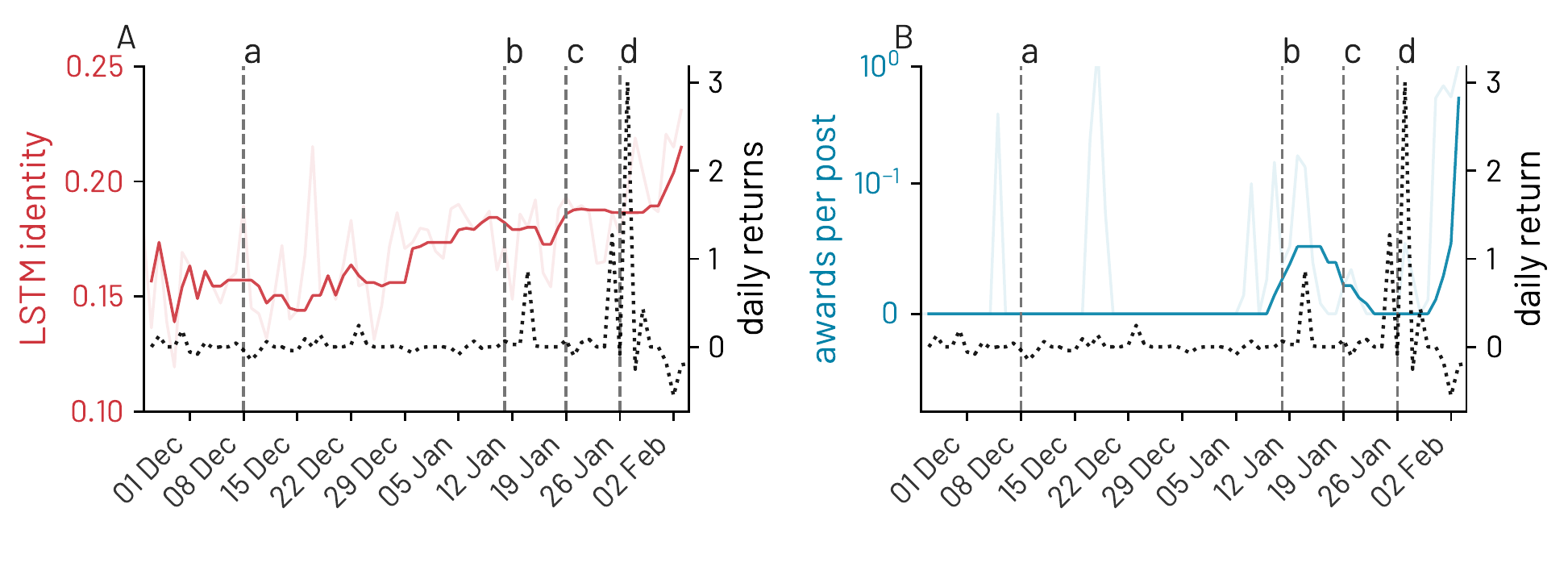}
		\caption{\emph{Evolution of group identity in time}. Group identity over time calculated as the fraction of messages that contain language expressions of identity according to a deep-learning classifier (a), and average number of Reddit awards that posts received at a given time (b). Raw signal is displayed together with a smoothed signal on a 7-days rolling median.}
		\label{fig:SIidentity}
	\end{figure*}
	
	Group identity is a broad concept that can be operationalized in multiple ways. To check the robustness of the results obtained with LIWC, we implemented two alternative measures of identity.
	
	The first is based on language and uses a readily-available classifier of identity expressions~\cite{choi20ten}, implemented with a Long Short-Term Memory neural network (LSTM)~\cite{hochreiter1997long}, a type of Recurrent Neural Network (RNN) that is commonly used in a variety of NLP tasks~\cite{sundermeyer2012lstm}.  Given in input a sentence, the classifier outputs a score in the range $[0,1]$ that expresses the likelihood that the message contains a verbal expression of identity. It is pre-trained on \num{9000} manually-labeled sentences from Reddit and it achieves an Area Under the ROC Curve (AUC) of $0.75$. AUC is a standard performance metric that assesses the ability of a classifier to rank positive and negative instances by their likelihood score, and gets the value of 1 for perfect probabilistic classification. To each message, we assign an identity score equal to the maximum score of its sentences. This approach reflects the theoretical interpretation of the use of social dimensions in language~\cite{deri18coloring}: a dimension is conveyed effectively through language even when expressed only briefly. Following the original approach~\cite{choi20ten}, we binarize the identity scores using a threshold equal to the value of the $90\textsuperscript{th}$ percentile of the empirical distribution of all the scores; such a high threshold favors precision over recall, thus focusing on a small set of messages that are very likely to contain expressions of identity. To measure group identity over time, we calculated the fraction of messages that the classifier marked as positive.
	
	As a second alternative measure of group identity, we used Reddit awards. The practice of awarding is rather frequent in \wsbs, and it requires a fee from the awarder (paid in virtual Reddit coins), thus representing a stronger endorsement than upvoting a post or commenting on it. The abundant use of awards has become a distinctive norm of \wsbs as a way to remark community identity~\cite{boylston2021wallstreetbets}. As a proxy of  non-verbal identity, we calculated the average number of awards received by submission within the time window considered.
	
	Figure~\ref{fig:SIidentity} shows that the frequency of linguistic expressions of identity increases slowly over time, concurrent with the activity of commits, and it soars considerably after the maximum peak of financial returns. The number of assigned awards per post increased dramatically after the peak of returns, likely as a signal of celebration of the community's success, but spiked also earlier in mid January, when the commitment activity started to increase considerably. Overall, these results strengthen the evidence that a process of identity construction has accompanied the dynamic of commitment.

	\subsection{Evolution of the discussion network in time}
	\label{sec:SInetworkEvolution}
	
\begin{figure*}
		\centering
		\includegraphics[width=\linewidth]{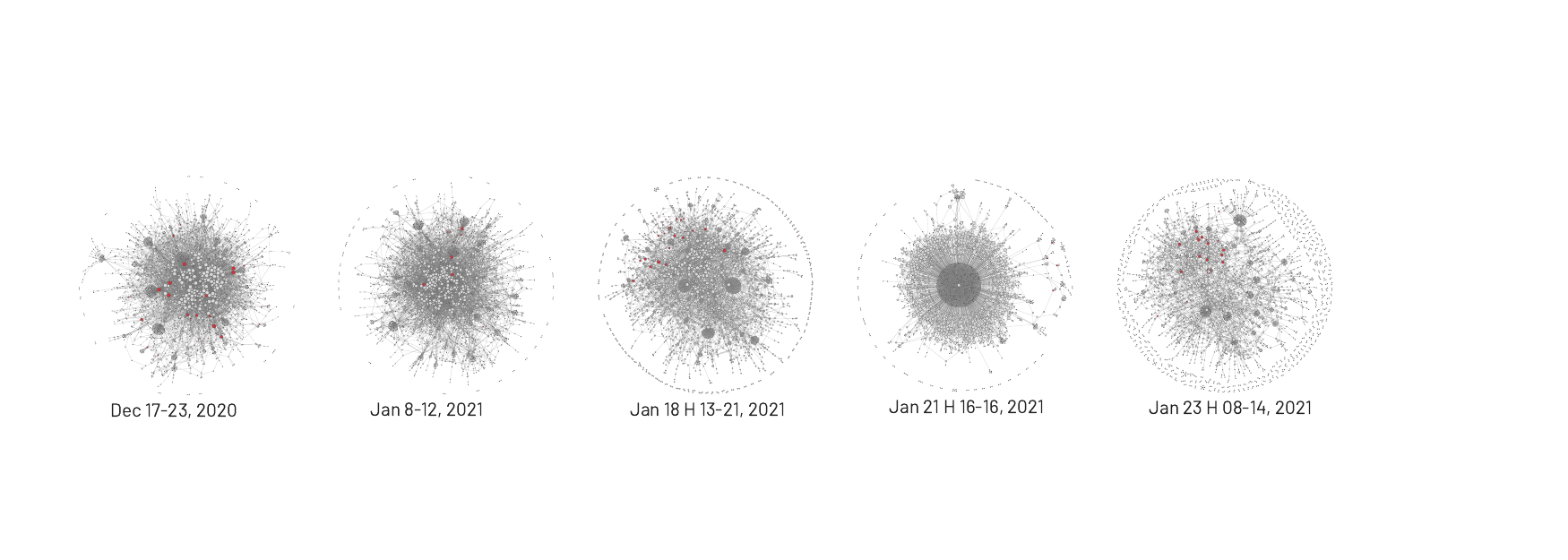}
		\caption{\emph{The evolution of the discussion networks.} Networks are built in such a way that each snapshot contains $3\,000$ different nodes. While before December 2020 involving thousands of authors in the discussion required several days of \wsbs activity, in late January 2021 thousands of users were interacting and discussing the topic simultaneously, reaching the $3\,000$ limit in less than one hour.}
		\label{fig:SInetworkevolution}
	\end{figure*}

	The discussion network about \gme evolves in time, as new posts and comments are continuously added. 
	Not only the network is dynamics, but the topological properties of static snapshots, i.e. the static networks reconstructed by data collected within fixed time intervals, change in time, reflecting the non-stationary evolution of the discussion structure.  This behavior is showed in Fig. 2 of the main text and more in detail in Figure~\ref{fig:SInetworkevolution}, which displays five different snapshot constructed by slicing the data in such a way that each network has a total of $3\,000$ unique authors interacting in the specific time window. 
	With this technique, the size of the windows changes depending on the frequency of posting. In particular, while in mid-December and at the beginning of January to build a network with $300$ different authors requires to aggregate within the same slice data from multiple days, after mid-January 2021 such a number of unique authors is interacting in few hours or less.
	
\begin{figure*}
		\centering
		\includegraphics[width=.45\linewidth]{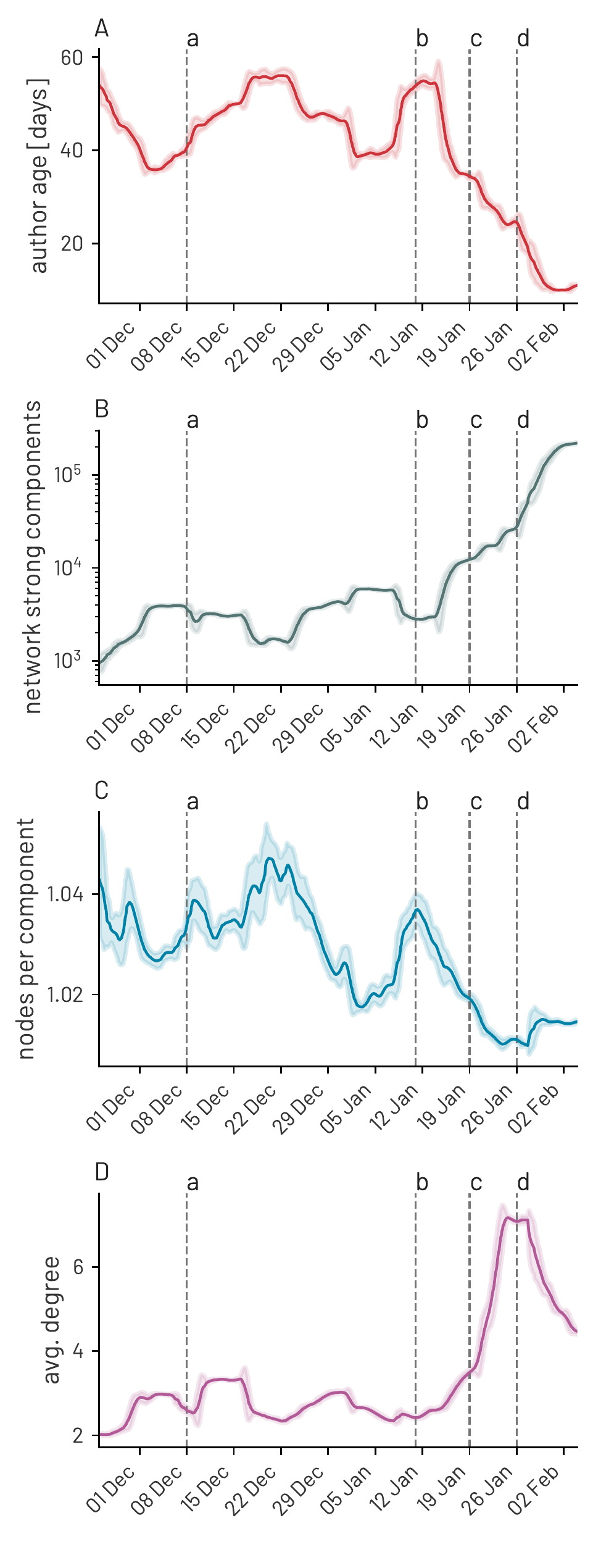}
		\caption{\emph{Topological properties of the discussion networks as a function of time.} (a) Average number of days since first submission for authors interacting within a $7$ day window. (b) Number of network components. (c) Number of nodes per component, excluding the giant component and its nodes. (d) Average degree of the networks.}
		\label{fig:SInetworkmetrics}
	\end{figure*}

	In the main text, we showed that the network's topology strongly changes in time, indicating that the structure of the discussion evolves towards a collective conversation. Fig.~\ref{fig:SInetworkmetrics} shows a few more features supporting this picture. In particular, in Fig.~\ref{fig:SInetworkmetrics}(a), we show how, as the process evolves, the age of users discussing on the subreddit decreases significantly. At the same time, Fig.~\ref{fig:SInetworkmetrics}(b) shows how the number of isolated components of the network constructed with windows of the duration rapidly increases after event ``b''. Figure~\ref{fig:SInetworkmetrics}(c) complements this finding, revealing that, as the number of components grow, the average number of nodes within the component decreases. Here we use as components the strongly connected components of each directed network constructed from a $7$ day window. The low average also reveal that the vast majority of components consists of only few nodes commenting a submission without ``out-of-comment-tree'' and no-reciprocal interactions. Figure~\ref{fig:SInetworkmetrics}(d) shows the evolution of the average node degree in time. Its increase starts right after the new directors announcement (event ``b'').

	\subsection{Robustness of commitment activity and reach with respect to the network's reconstruction}\label{sec:SIslicingRobustness}
	
	The way in which the discussion network is reconstructed is determinant to measure the distribution of commitment activity and its reach on the network.  
	While in the main text we showed results for a temporal slicing (networks are constructed slicing data within a 7-days window), here we present the results for two additional slicing strategies: windows of data with a constant number of unique authors or submissions.
	
	\begin{figure*}
		\centering
		\includegraphics[width=\linewidth]{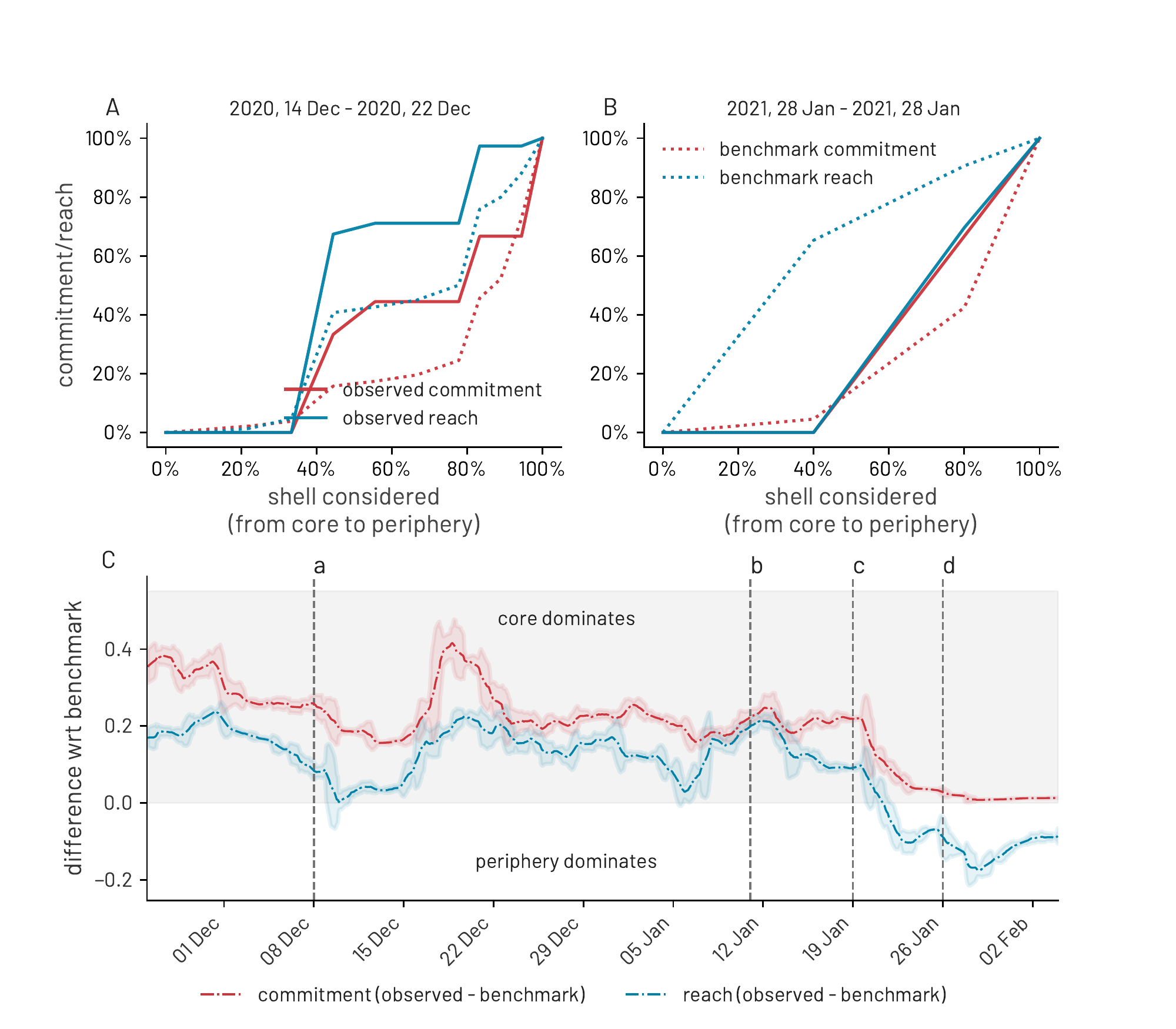}
		\caption{\emph{Evolution of commitment and reach on networks with $3\,000$ nodes per network slice. Equivalent to Figure 3 of the main text.}}
		\label{fig:SInodeSlicing}
	\end{figure*}
	Figure SI~\ref{fig:SInodeSlicing} shows the evolution of commitment and reach on networks with $3\,000$ nodes per slice. Imposing each slice to have a fixed number of authors is equivalent to requiring that each network slice has a fixed number of nodes. Importantly, we stress that the slicing is pruning links between authors which are activated at times $t< t_{start}$ or $t\ge t_{end}$, where $t_{start}$ is the lower time-bound of the slice, while $t_{end}$ is the upper time-bound of the slice, i.e. the time at which the $3\,001th$ author is activating a new link. 
	Panel (a) and (b) shows the cumulative distribution of commitment and reach (red and blue respectively) as a function of node coreness (from more central nodes to the peripheral ones). Two slices are selected from two different time windows, one at mid-December 2020 and the second during January 28,2021. Panel (c) shows a significant reduction in the commitment and reach from core nodes also with this slicing strategy.
	
	\begin{figure*}
		\centering
		\includegraphics[width=\linewidth]{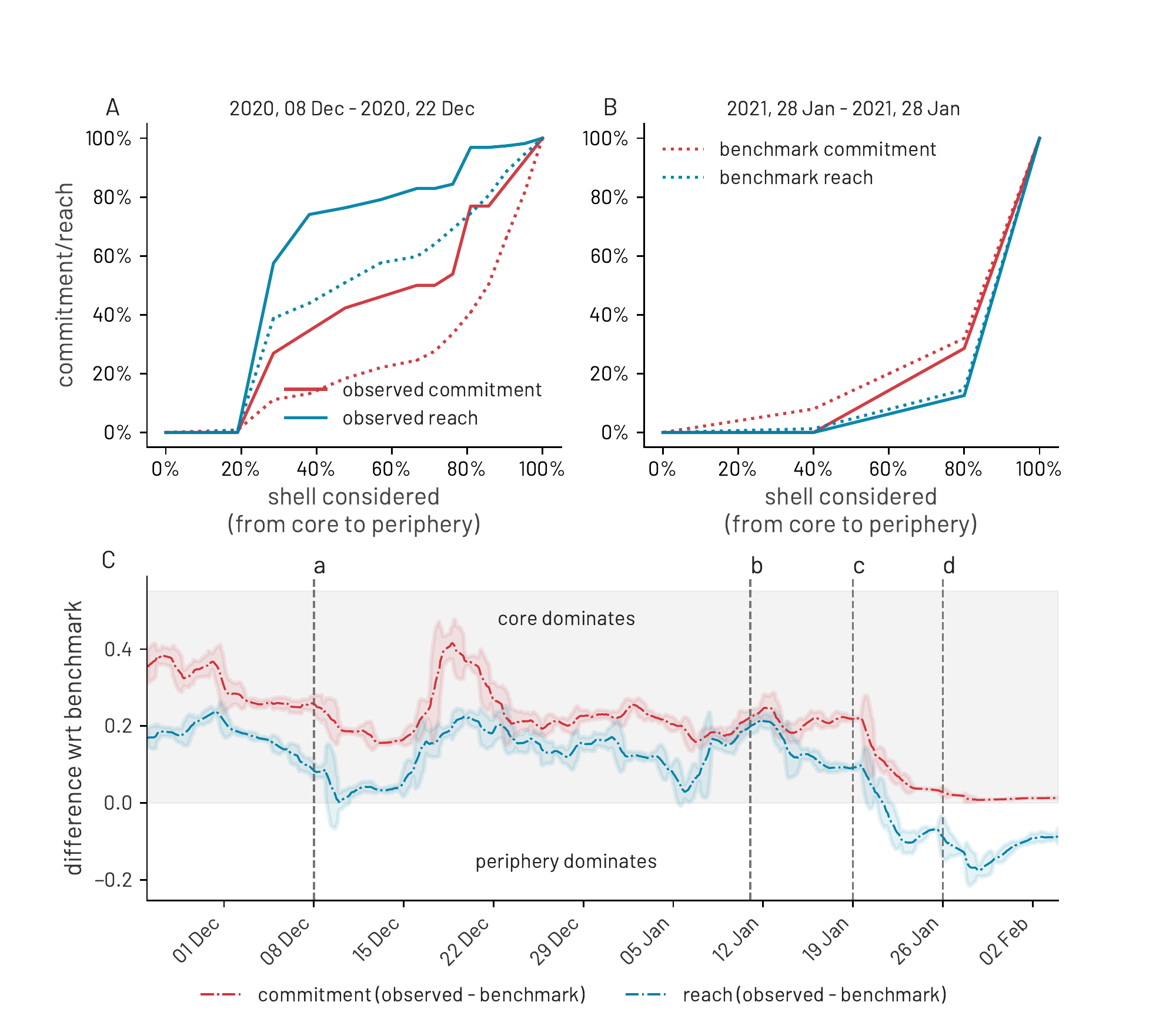}
		\caption{\emph{Evolution of commitment and reach on networks with $500$ submissions per network slice. Slicing is performed rolling the window including each time the next $25$ posts submitted and removing the oldest $25$. Equivalent to Figure 3 of the main text.}}
		\label{fig:SIsubsSlicing}
	\end{figure*}
	Figure SI~\ref{fig:SIsubsSlicing} shows the evolution of commitment and reach on networks constructed including all posts and comments within the temporal boundaries delimited by the oldest and latest time of a consecutive set of $500$ posts. In this framework while $t_{start}$ is the time at which the oldest post was submitted, $t_{end}$ is the time at which the $501th$ post is submitted. Also in this case, all links activated outside these temporal boundaries are neglected from the analysis. Also this slicing technique shows a significant reduction in the commitment and reach from core nodes, confirming the results discussed in the main text.
	
\end{document}